\documentclass[apj]{emulateapj}
\usepackage{apjfonts}
\usepackage{url}
\usepackage{rotating}

\slugcomment{Accepted for publication in the Astrophysical Journal}

\shorttitle{Quiescent galaxies at $z\sim2$}
\shortauthors{Szomoru et al.}

\begin{document}
\title{Sizes and surface brightness profiles of quiescent galaxies at
  $z\sim 2$.}

\author{
Daniel Szomoru\altaffilmark{1},
Marijn Franx\altaffilmark{1},
Pieter~G.~van Dokkum\altaffilmark{2}
}

\altaffiltext{1}
{Leiden Observatory, Leiden University, P.O. Box 9513, 2300 RA Leiden, The Netherlands}
\altaffiltext{2}
{Department of Astronomy, Yale University, New Haven, CT 06520-8101, USA}

\begin{abstract}
We use deep \textit{Hubble Space Telescope} Wide Field Camera 3
near-infrared imaging obtained of the GOODS-South field as part of the
CANDELS survey to investigate a stellar mass-limited sample of
quiescent galaxies at $1.5 < z < 2.5$. We measure surface brightness
profiles for these galaxies using a method that properly measures low
surface brightness flux at large radii. We find that quiescent
galaxies at $z\sim2$ very closely follow S\'ersic profiles, with
$n_{median} = 3.7$, and have no excess flux at large radii. Their
effective radii are a factor $\sim4$ smaller than those of
low-redshift quiescent galaxies of similar mass. However, there is
significant spread in sizes ($\sigma_{\log_{10} r_e} = 0.24$), with
the largest $z\sim2$ galaxies lying close to the $z=0$ mass-size
relation. We compare the stellar mass surface density profiles with
those of massive elliptical galaxies in the Virgo cluster and confirm
that most of the mass-growth which occurs between $z\sim2$ and $z=0$
must be due to accretion of material onto the outer regions of the
galaxies. Additionally, we investigate the evolution in the size
distribution of massive quiescent galaxies. We find that the minimum
size growth required for $z\sim2$ quiescent galaxies to fall within
the $z=0$ size distribution is a factor $\sim2$ smaller than the total
median size growth between $z\sim2$ and $z=0$.
\end{abstract}


\section{Introduction}
Quiescent galaxies make up a considerable fraction of the massive
galaxy population at $z=2$ (e.g., \citealt{fra03}; \citealt{dad05};
\citealt{kri06}). Their structural evolution has been the subject of
considerable discussion, focusing in particular on their extremely
compact nature compared to low redshift galaxies of similar mass
(e.g., \citealt{dad05}; \citealt{tru06}; \citealt{tof07};
\citealt{wel08}; \citealt{dok08}; \citealt{dam09}; \citealt{hop09};
\citealt{sar09}; \citealt{dok09b}; \citealt{cas10}; \citealt{man10};
\citealt{cas11}). The early formation and subsequent evolution of
these massive, compact objects presents a considerable challenge to
current models of galaxy formation and evolution (e.g.,
\citealt{wuy10}; \citealt{ose11}). It is unclear what the structure of
the progenitors of these galaxies is, and the lack of extremely
compact massive galaxies at low redshift implies considerable size
evolution between $z=2$ and $z=0$ (\citealt{tru09};
\citealt{tay10a}). However, efforts to accurately quantify this
evolution are hindered by uncertainties. The apparent compactness of
$z\sim2$ quiescent galaxies may simply be an observational effect:
photometric masses may be systematically overestimated due to modeling
uncertainties, and sizes may be underestimated due to a lack of
imaging depth (\citealt{hop09}; \citealt{muz09}).

Due to the difficulty of obtaining high-quality spectra of quiescent
galaxies at $z > 1.5$, dynamical masses have only been measured for a
few such galaxies (\citealt{cap09}; \citealt{cen09}; \citealt{dok09b};
\citealt{ono10}; \citealt{san11}). Instead, photometric stellar masses
are used, which are subject to considerable uncertainties due to e.g.,
the quality of the stellar libraries used in modeling the spectral
energy distribution (SED), or incorrect assumptions about the shape of
the initial mass function (IMF). These uncertainties can result in
systematic errors of up to a factor $\sim6$ \citep{con09}. At low
redshift there is good agreement between stellar masses determined by
photometric SED fitting methods and dynamical masses
(\citealt{tay10b}). Whether this is also the case at high redshift is
unclear (e.g., \citealt{san11}; \citealt{bez11}; \citealt{mar11}).

The second large source of uncertainty lies in the size determination
of these galaxies. The compact objects observed at $z\sim2$ may be
surrounded by faint extended envelopes of material, which could be
undetected by all but the deepest data. Stacking studies have been
used to obtain constraints on the average surface brightness profile
of compact galaxies (e.g., \citealt{wel08}; \citealt{dok08};
\citealt{cas10}). However, detailed analysis of individual galaxies is
more difficult, primarily due to the limited number of compact
galaxies for which ultradeep near-infrared (NIR) data are
available. \cite{szo10} carried out an analysis on a $z=1.91$ compact
quiescent galaxy in the Hubble Ultra Deep Field (HUDF) and confirmed
its small size.

In this Paper we expand the analysis of \cite{szo10} using a stellar
mass-limited sample of 21 quiescent galaxies. We make use of deep
\textit{Hubble Space Telescope} Wide Field Camera 3 (\textit{HST}
WFC3) data from the CANDELS GOODS-South observations to investigate
the surface brightness profiles of quiescent galaxies at
$z\sim2$. These observations are not as deep as the HUDF data, but
cover a much larger area, allowing us to study a statistically more
meaningful sample. We measure the surface brightness profile of each
individual galaxy and investigate deviations from S\'ersic
profiles. Additionally, we compare the size distribution and profile
shapes of $z\sim2$ galaxies to those of low redshift quiescent
galaxies. Throughout the Paper, we assume a $\Lambda$CDM cosmology
with $\Omega_m = 0.3$, $\Omega_\Lambda = 0.7$ and $H_0 = 70$ km
s$^{-1}$ Mpc$^{-1}$. All stellar masses are derived assuming a Kroupa
IMF \citep{kro01}. All effective radii are circularized and magnitudes
are in the AB system.

\begin{figure*}
\epsscale{1.1}
\plotone{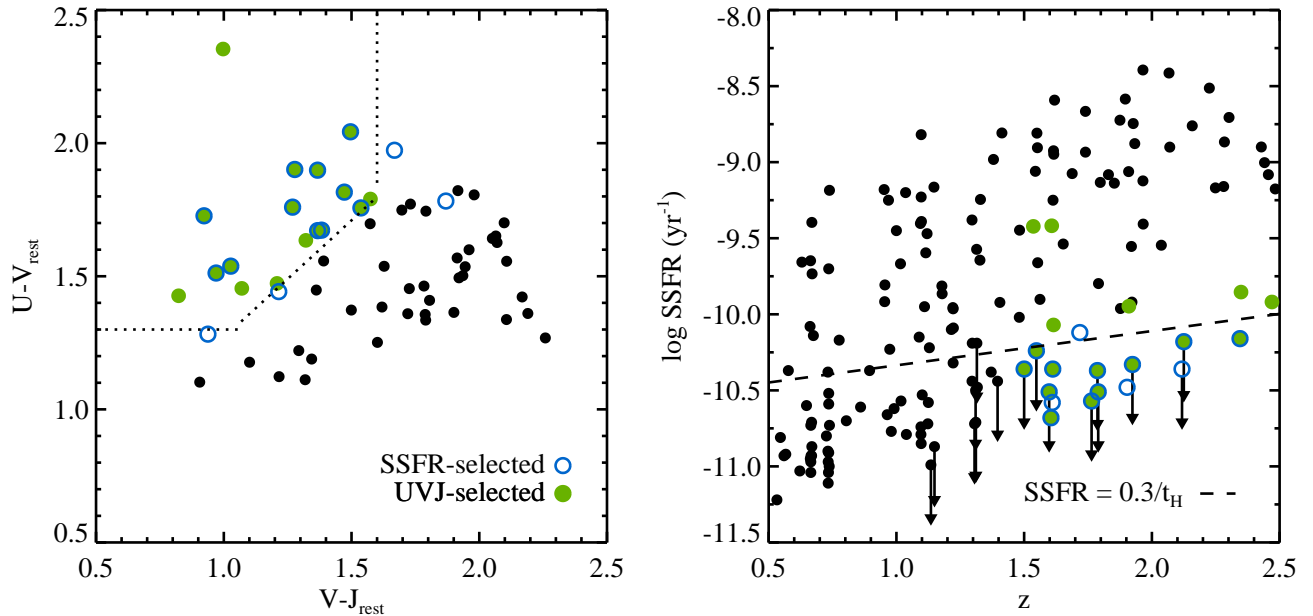}
\caption{Left panel: rest-frame $U-V$ and $V-J$ colors of galaxies in
  the CANDELS GOODS-South deep field at $1.5 < z < 2.5$. Right panel:
  specific star formation rates as a function of redshift. Arrows
  indicate upper limits. The dashed line indicates where the specific
  star formation rate is equal to $0.3/t_H$. Quiescent galaxies
  selected using the $UVJ$ color criterion are shown as filled green
  circles. Galaxies which are selected as quiescent based on their
  SSFRs are shown as open blue circles. There is good agreement
  between the two selection criteria. Both the $UVJ$-selected galaxies
  and the SSFR-selected galaxies are included in our quiescent galaxy
  sample.}\label{fig:uvj}
\end{figure*}

\section{Data and sample selection}

\begin{figure*}
\epsscale{1.1}
\plotone{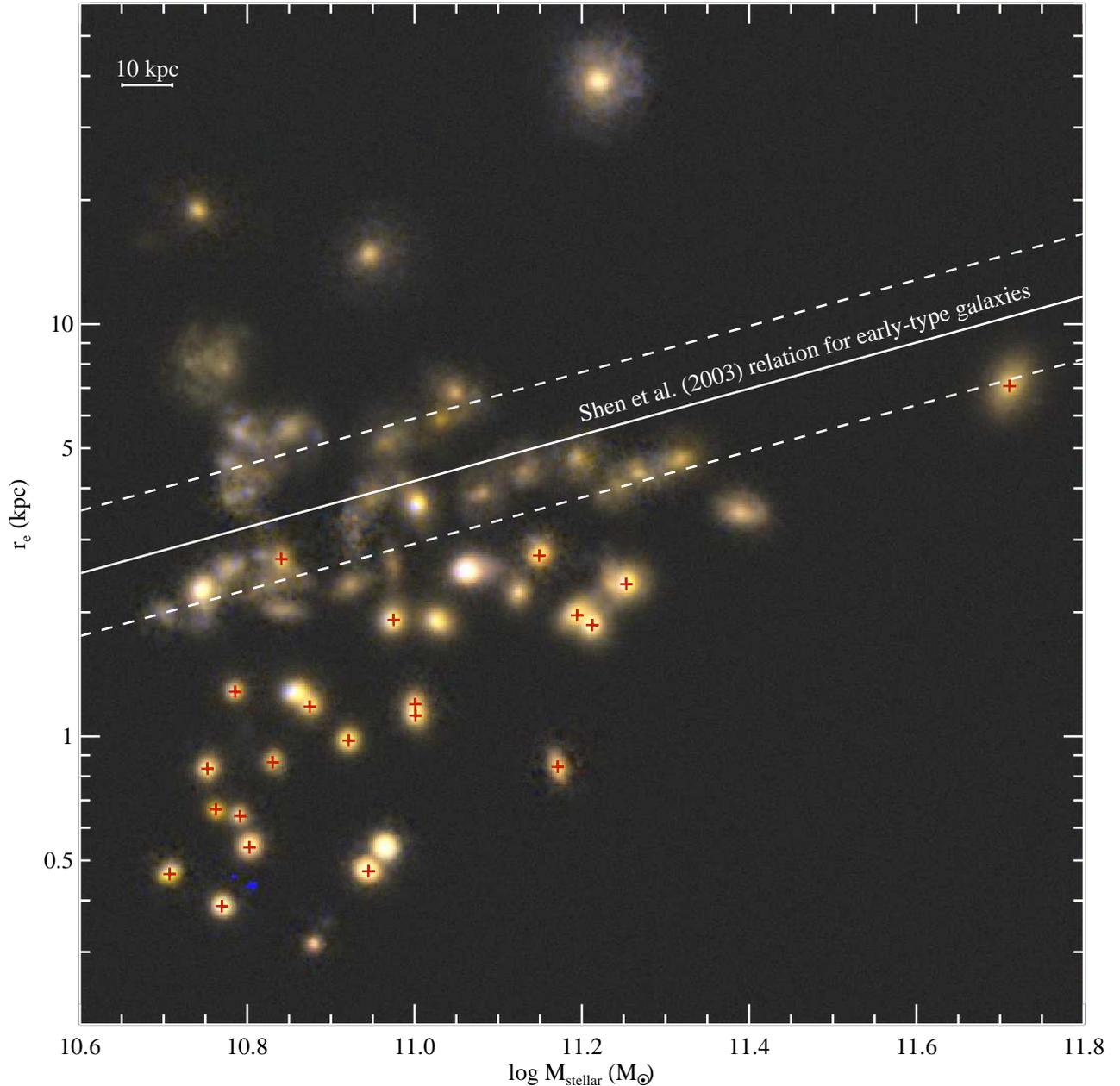}
\caption{Stellar masses and sizes of galaxies at $1.5 < z < 2.5$ with
  $M_{stellar} > 5\times10^{10} M_\odot$. Color images are composed of
  rest-frame $U_{336}$, $B_{438}$ and $g_{475}$ images, obtained from
  observed $I_{814}$, $J_{125}$ and $H_{160}$ images. Galaxies which
  are included in our quiescent sample are indicated with red
  crosses. Although we do not select based on morphology, almost all
  galaxies in our quiescent sample are compact, bulge-dominated, and
  have red colors.}\label{fig:cutouts}
\end{figure*}

We use NIR data taken with \textit{HST} WFC3 as part of the CANDELS
survey (\citealt{gro11}; \citealt{koe11}). This survey will target
approximately 700 square arcminutes to 2 orbit depth in $Y_{105}$,
$J_{125}$ and $H_{160}$ (COSMOS, EGS and UDS fields), as well as
$\sim120$ square arcminutes to 12 orbit depth (GOODS-South and
GOODS-North fields). These NIR observations are complemented with
parallel \textit{HST} ACS exposures in $V_{606}$ and $I_{814}$. We use
the deepest publicly available data, reduced by \cite{koe11}, which
consist of $I_{814}$, $J_{125}$ and $H_{160}$ observations to 4-orbit
depth of a $\sim60$ square arcminute section of the GOODS-South
field. The full width at half-maximum of the point-spread function
(PSF) is $\approx 0.18$ arcsec for the WFC3 observations and $\approx
0.11$ arcsec for the ACS observations. The images have been drizzled
to a pixel size of 0.06 arcsec for the WFC3 observations and 0.03
arcsec for the ACS observations (see \cite{koe11} for details).

Galaxies are selected in the GOODS-South field using the
$K_s$-selected FIREWORKS catalog \citep{wuy08}. This catalog combines
observations of the Chandra Deep Field South ranging from ground-based
$U$-band data to \textit{Spitzer} 24 $\mu$m data, and includes
spectroscopic redshifts where available, as well as photometric
redshifts derived using EAZY \citep{bra08}. These photometric
redshifts have a median $\Delta z/(1+z) = -0.001$ with a normalized
median absolute deviation of $\sigma_\mathrm{NMAD} = 0.032$
\citep{wuy08}. Stellar masses were estimated from SED fits to the full
photometric data set (N. M. F\"orster Schreiber et al. 2011, in
preparation), assuming a Kroupa IMF and the stellar population models
of \cite{bru03}.

We select all galaxies with $1.5 < z < 2.5$ and stellar masses above
$5\times10^{10} M_\odot$, which is the completeness limit in this
redshift range \citep{wuy09}. In order to ensure that we include all
quiescent galaxies we explore both a color-color selection (the $UVJ$
selection described in \cite{wil09}) and a selection based on specific
star formation rate (SSFR). In the left panel of Figure~\ref{fig:uvj}
we show the rest-frame $U-V$ and $V-J$ colors of all $z\sim2$ galaxies
in the field. The dashed lines indicate the quiescent galaxy selection
limits from \cite{wil09}. Galaxies which fall within the dashed lines
(green dots) have SEDs that are consistent with red, quiescent
galaxies. \cite{pat11} have shown that this selection method is very
effective at separating dust-reddened starforming galaxies from truly
quiescent galaxies. As an alternative to the UVJ selection we also
select galaxies based on their SSFR. In the right-hand panel of
Figure~\ref{fig:uvj} we show the SSFRs of galaxies as a function of
redshift. The SSFRs are estimated from the UV and 24 $\mu$m fluxes, as
discussed in \cite{wuy09}. The dashed line shows our selection limit,
below which the SSFR is lower than $0.3/t_H$, where $t_H$ is the
Hubble time. There is generally very good agreement between the two
selection criteria, although several galaxies that seem to be
quiescent based on their SSFRs are not selected by the $UVJ$ method,
and vice versa. We find no significant difference in the distribution
of structural parameters of galaxies selected by either method; the
median values are equal to within 6 percent, for the effective radii,
S\'ersic indices and axis ratios. This is expected, given the large
overlap between the two samples. Since we wish to be as complete as
possible we combine the two selection methods and include all galaxies
selected by either method. This results in a sample of 21 quiescent
galaxies, whose properties are summarized in Table~\ref{tab:table}.

To illustrate the effects of our selection on galaxy morphology we
show color images of all galaxies with $1.5 < z < 2.5$ and
$M_{stellar} > 5\times10^{10} M_\odot$ in the stellar mass-size plane
in Figure~\ref{fig:cutouts}. The color images are constructed from
PSF-matched rest-frame $U_{336}$, $B_{438}$ and $g_{475}$ images,
obtained by interpolating between the observed $I_{814}$, $J_{125}$
and $H_{160}$ images. Although we do not select based on morphology,
the galaxies in our quiescent sample (indicated with red crosses) are
generally very compact, bulge-dominated systems with relatively red
colors. Interestingly, all starforming systems at $z\sim2$ appear to
have a well-defined red core, as was also pointed out by \cite{szo11}
(but also see, e.g., \citealt{for11a}; \citealt{for11b}).

\begin{sidewaystable*}
\centering
\caption{Galaxy properties.\label{tab:table}}
\begin{tabular}{l l c c c c c c c c c c r}
\tabletypesize{\scriptsize}
\\
\hline \hline
\colhead{ID$^\mathrm{a}$} & \colhead{$z$} & \colhead{R.A} & \colhead{Dec} & \colhead{$M_\mathrm{stellar}^\mathrm{b}$} & \colhead{SSFR} & \colhead{$U-V_\mathrm{rest}$} & \colhead{$V-J_\mathrm{rest}$} & \colhead{mag$_{H,app}^\mathrm{c}$} & \colhead{$r_e^\mathrm{c}$} & \colhead{$n^\mathrm{c}$} & \colhead{b/a} & \colhead{P.A.$^\mathrm{d}$} \\
\colhead{} & \colhead{} & \colhead{} & \colhead{} & \colhead{($\log M_\odot$)} & \colhead{($\log$ yr$^{-1}$)} & \colhead{} & \colhead{} & \colhead{(AB)} & \colhead{(kpc)} & \colhead{} & \colhead{} & \colhead{(deg)} \\
\hline
1060 & 2.345$^*$ & 53.069829 & -27.880467 & 11.14 & -10.16 & 1.75 & 0.92 & $22.21 \pm 0.05$ & $2.75 \pm 1.60$ & $ 9.21 \pm 1.10$ & $0.70 \pm 0.01$ & $-45.2 \pm  1.6$ \\
1088 & 1.752$^*$ & 53.065570 & -27.878805 & 10.75 & -10.12 & 1.28 & 0.94 & $21.84 \pm 0.03$ & $0.83 \pm 0.11$ & $ 5.50 \pm 0.67$ & $0.87 \pm 0.02$ & $-62.5 \pm 13.1$ \\
1289 & 1.759$^*$ & 53.116186 & -27.871904 & 11.00 & -10.51 & 2.04 & 1.50 & $22.35 \pm 0.02$ & $1.20 \pm 0.22$ & $ 3.26 \pm 0.40$ & $0.58 \pm 0.01$ & $-8.5  \pm  0.8$ \\
1831 & 1.536     & 53.076366 & -27.848700 & 11.25 &  -9.42 & 1.47 & 1.21 & $20.71 \pm 0.02$ & $2.34 \pm 0.33$ & $ 3.68 \pm 0.16$ & $0.92 \pm 0.01$ & $-44.6 \pm  3.2$ \\
1971 & 1.608     & 53.150661 & -27.843604 & 10.84 &  -9.42 & 1.63 & 1.32 & $21.71 \pm 0.03$ & $2.69 \pm 0.79$ & $ 5.07 \pm 0.31$ & $0.87 \pm 0.01$ & $27.0  \pm  2.1$ \\
2227 & 1.612     & 53.150165 & -27.834522 & 10.98 & -10.36 & 1.54 & 1.03 & $21.40 \pm 0.02$ & $1.92 \pm 0.26$ & $ 3.76 \pm 0.15$ & $0.84 \pm 0.01$ & $-6.5  \pm  2.1$ \\
2514 & 1.548$^*$ & 53.151413 & -27.825886 & 10.79 & -10.24 & 1.67 & 1.38 & $21.96 \pm 0.06$ & $1.28 \pm 0.29$ & $ 5.73 \pm 0.93$ & $0.86 \pm 0.03$ & $7.0   \pm  5.9$ \\
2531 & 1.598$^*$ & 53.171735 & -27.825672 & 10.87 & -10.51 & 1.90 & 1.28 & $21.93 \pm 0.02$ & $1.18 \pm 0.12$ & $ 4.08 \pm 0.30$ & $0.95 \pm 0.02$ & $7.6   \pm 11.6$ \\
2856 & 1.759$^*$ & 53.216633 & -27.814310 & 10.83 & -10.37 & 1.76 & 1.54 & $22.90 \pm 0.01$ & $0.87 \pm 0.03$ & $ 1.20 \pm 0.08$ & $0.63 \pm 0.02$ & $-16.5 \pm  1.0$ \\
2993 & 2.470     & 53.163233 & -27.808962 & 10.71 &  -9.92 & 2.35 & 1.00 & $23.38 \pm 0.02$ & $0.46 \pm 0.04$ & $ 1.01 \pm 0.09$ & $0.32 \pm 0.04$ & $-63.1 \pm  1.0$ \\
3046 & 2.125$^*$ & 53.116519 & -27.806731 & 10.80 & -10.18 & 1.51 & 0.97 & $22.22 \pm 0.02$ & $0.54 \pm 0.02$ & $ 3.59 \pm 0.34$ & $0.70 \pm 0.02$ & $-45.8 \pm  3.0$ \\
3119 & 2.349     & 53.123107 & -27.803355 & 10.94 &  -9.85 & 1.43 & 0.82 & $21.97 \pm 0.03$ & $0.47 \pm 0.06$ & $ 5.09 \pm 0.60$ & $0.49 \pm 0.04$ & $79.2  \pm  1.5$ \\
3242 & 1.910     & 53.158831 & -27.797119 & 10.77 &  -9.95 & 1.45 & 1.07 & $22.10 \pm 0.02$ & $0.39 \pm 0.03$ & $ 4.17 \pm 0.45$ & $0.62 \pm 0.03$ & $57.1  \pm  3.0$ \\
3548 & 1.500$^*$ & 53.202356 & -27.785436 & 10.76 & -10.36 & 1.67 & 1.37 & $22.40 \pm 0.03$ & $0.67 \pm 0.05$ & $ 3.75 \pm 0.48$ & $0.65 \pm 0.04$ & $55.5  \pm  2.4$ \\
3829 & 1.924$^*$ & 53.069966 & -27.768143 & 10.79 & -10.33 & 1.90 & 1.37 & $22.85 \pm 0.05$ & $0.64 \pm 0.14$ & $ 4.24 \pm 1.15$ & $0.66 \pm 0.03$ & $44.7  \pm  3.2$ \\
4850 & 2.118$^*$ & 53.012891 & -27.705730 & 11.17 & -10.36 & 1.97 & 1.67 & $22.68 \pm 0.02$ & $0.84 \pm 0.09$ & $ 2.72 \pm 0.42$ & $0.20 \pm 0.02$ & $17.0  \pm  0.3$ \\
5890 & 1.756$^*$ & 53.174620 & -27.753362 & 10.92 & -10.57 & 1.76 & 1.27 & $22.09 \pm 0.01$ & $0.98 \pm 0.04$ & $ 1.89 \pm 0.12$ & $0.92 \pm 0.02$ & $-14.6 \pm  3.6$ \\
6097 & 1.903     & 53.140997 & -27.766706 & 11.21 & -10.48 & 1.44 & 1.22 & $21.30 \pm 0.03$ & $1.86 \pm 0.43$ & $ 5.26 \pm 0.56$ & $0.79 \pm 0.02$ & $2.8   \pm  2.2$ \\
6187 & 1.610     & 53.044923 & -27.774363 & 11.71 & -10.58 & 1.76 & 1.87 & $20.37 \pm 0.01$ & $7.08 \pm 1.30$ & $ 2.77 \pm 0.05$ & $0.61 \pm 0.01$ & $-30.1 \pm  0.2$ \\
6194 & 1.605     & 53.052217 & -27.774766 & 11.19 & -10.68 & 1.82 & 1.47 & $21.18 \pm 0.02$ & $1.97 \pm 0.14$ & $ 2.04 \pm 0.06$ & $0.57 \pm 0.01$ & $-56.3 \pm  0.4$ \\
6246 & 1.615     & 53.043813 & -27.774666 & 11.00 & -10.07 & 1.76 & 1.57 & $21.71 \pm 0.03$ & $1.12 \pm 0.16$ & $10.10 \pm 2.38$ & $0.64 \pm 0.02$ & $-19.6 \pm  1.5$ \\
\hline \\
\multicolumn{13}{l}{$^\mathrm{a}$FIREWORKS ID \citep{wuy08}} \\
\multicolumn{13}{l}{$^\mathrm{b}$Masses are corrected to account for the difference between the catalog magnitude and our measured magnitude.} \\
\multicolumn{13}{l}{$^\mathrm{c}$Magnitudes, effective radii and S\'ersic indices are derived from the $H_{160}$ band residual-corrected profiles discussed in Section~\ref{sec:profiles}.} \\
\multicolumn{13}{l}{$^\mathrm{d}$Position angles are measured counterclockwise with respect to North.} \\
\multicolumn{13}{l}{$^\mathrm{*}$No spectroscopic redshifts are available for these galaxies; photometric redshifts are listed instead}
\end{tabular}
\end{sidewaystable*}

\section{Measuring surface brightness profiles}

Obtaining surface brightness profiles of high-redshift galaxies is
difficult, in large part due to the small size of these galaxies
compared to the PSF. Direct deconvolution of the observed images is
subject to large uncertainties. A common approach is therefore to fit
two-dimensional models, convolved with a PSF, to the observed
images. \cite{ser68} profiles are commonly used, since these have been
shown to closely match the surface brightness profiles of nearby
early-type galaxies (e.g., \citealt{cao93}; \citealt{gra03};
\citealt{tru04}; \citealt{fer06}; \citealt{cot07};
\citealt{kor09}). However, there is no reason that high-redshift
galaxies should exactly follow S\'ersic profiles.

An obvious way to account for deviations from a S\'ersic profile is by
using double-component fits, in which the deviations are approximated
by a second S\'ersic profile. Although this provides a closer
approximation to the true surface brightness profile than a
one-component fit, it still depends on assumptions regarding the shape
of the profile. We therefore use a technique which is more robust to
deviations from the assumed model and accurately recovers the true
intrinsic profile. This technique was first used in \cite{szo10}; we
summarize it here. First, we use the GALFIT package \citep{pen02} to
perform a conventional two-dimensional S\'ersic profile fit to the
observed image. For PSFs we use unsaturated stars brighter than $K =
22.86$ that are not contaminated by nearby sources. We verify the
quality of our stellar PSFs by comparing their radial profiles to each
other, and find that the profiles show small variations in half-light
radius of order $\sim2\%$. We find no systematic dependence of these
variations with magnitude. In order to estimate the effects of PSF
variations on our derived parameters we fit every galaxy using each of
the stars separately. We find that the derived total magnitudes, sizes
and S\'ersic indices vary by about 0.1\%, 3\% and 7\%, respectively.

After fitting a S\'ersic model profile we measure the residual flux
profile from the residual image, which is the difference between the
observed image and the best-fit PSF-convolved model. This is done
along concentric ellipses which follow the geometry of the best-fit
S\'ersic model. The residual flux profile is then added to the
best-fit S\'ersic profile, effectively providing a first-order
correction to the profile at those locations where the assumed model
does not accurately describe the data. The effective radius is then
calculated by integrating the residual-corrected profile out to a
radius of approximately 12 arcseconds ($\sim100$ kpc at $z\sim2$). We
note that the residual flux profile is not deconvolved for PSF;
however, we show below that this does not strongly affect the accuracy
of this method.

Errors in the sky background estimate are the dominant source of
uncertainty when deriving surface brightness profiles of faint
galaxies to large radii. Using the wrong sky value can result in
systematic effects. GALFIT provides an estimate of the sky background
during fitting. To ensure that this estimate is correct we inspect the
residual flux profile of each galaxy at radii between 5 and 15 arcsec
(approximately 40 to 120 kpc at $z=2$). Using this portion of the
residual flux profile we derive a new sky value and adjust the
intensity profile accordingly. We use the difference between the
minimum and maximum values of the residual flux profile within this
range of radii as an estimate of the uncertainty in the sky
determination.

\begin{figure*}
\epsscale{1.1}
\plotone{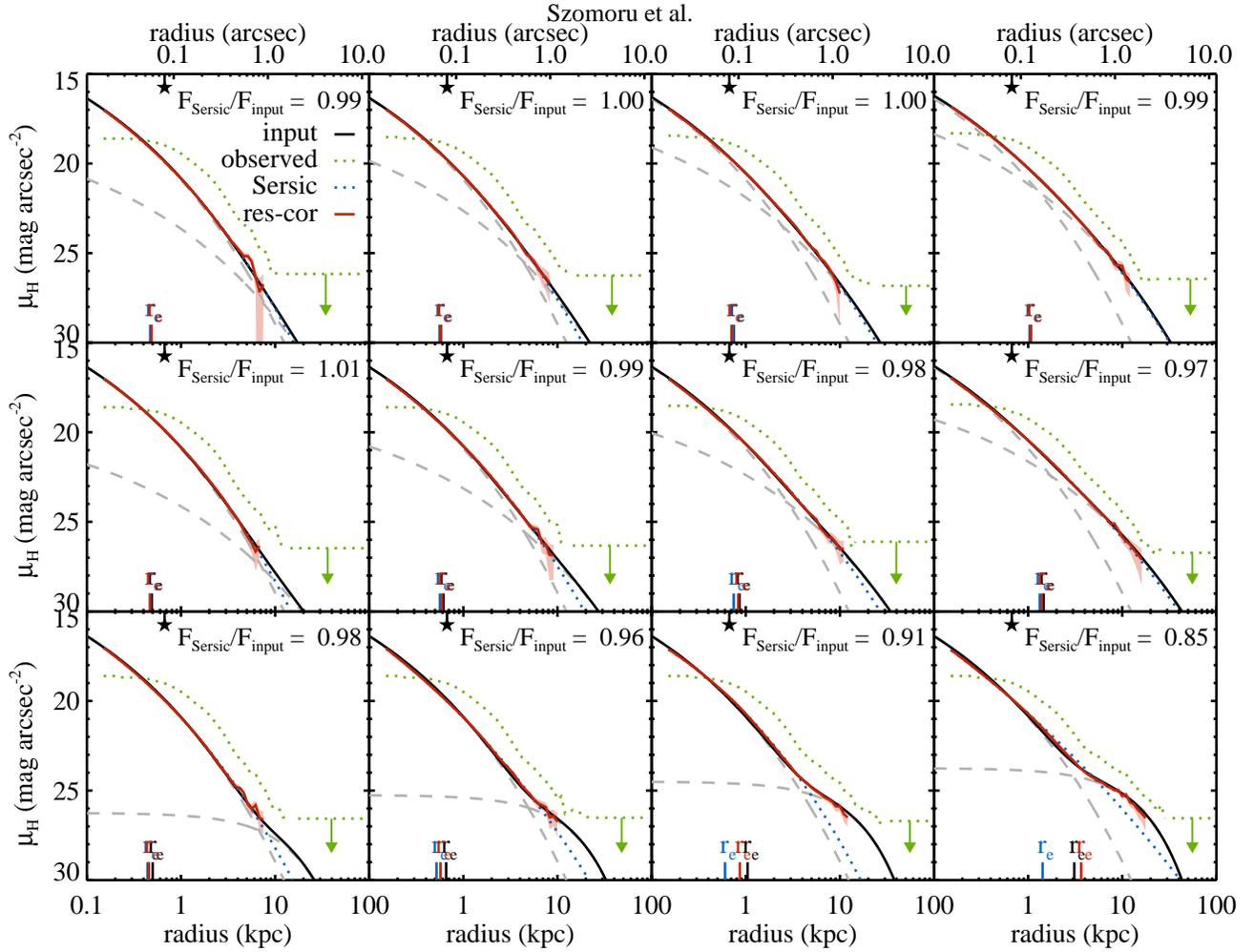}
\caption{Effectiveness of the residual-correction for recovering
  surface brightness profiles. The method was tested on a large number
  of simulated galaxies, composed of two components: one compact
  bright component, and an extended fainter component. A small
  selection is shown here. The input profiles are shown in black, with
  the dashed grey lines indicating the two subcomponents. The
  PSF-convolved ``observed'' profiles are shown in green. Direct
  S\'ersic fits are shown in blue, and the residual-corrected profiles
  are overplotted in red. The shaded light red regions indicate the
  1-$\sigma$ errors due to uncertainty in the sky estimation. The size
  of the PSF half width at half maximum (HWHM) is indicated on the top
  axis of each panel. Input effective radii are indicated in black on
  the bottom axes. Effective radii derived from the direct S\'ersic
  fits and from the residual-corrected profiles are indicated in blue
  and red, respectively. The fraction of the input flux within 10 kpc
  recovered by the Sersic fits $F_\mathrm{Sersic}/F_\mathrm{input}$ is
  given in each panel. The residual-corrected profiles clearly
  reproduce the input profiles more accurately than the simple
  S\'ersic fits, especially at large radii.}
\label{fig:test}
\end{figure*}

In \cite{szo10} this procedure was tested using simulated galaxies
inserted into \textit{HST} WFC3 data of the HUDF. Since the data used
in this Paper are shallower we have performed new tests. We create
images of simulated galaxies that consist of two components: one
compact elliptical component and a larger, fainter component that
ranges from disk-like to elliptical. The axis ratio and position angle
of the second component are varied, as are its effective radius and
total magnitude. The simulated galaxies are convolved with a PSF
(obtained from the data) and are placed in empty areas of the observed
$H_{160}$ band image. We then run the procedure described above to
extract surface brightness profiles and compare them to the input
profiles.

A selection of these simulated profiles is shown in
Figure~\ref{fig:test}. The input profiles are shown as solid black
lines. The dashed grey lines indicate the two subcomponents of each
simulated galaxy. The directly measured profiles are shown in
green. The best-fit S\'ersic models are shown in blue, and the
residual-corrected profiles are shown in red. The residual-corrected
profiles are plotted up to the radius where the uncertainty in the sky
determination becomes significant. The effectiveness of the
residual-correction method is clear: whereas a simple S\'ersic fit in
many cases under- or overpredicts the flux at $r > 5$ kpc, the
residual-corrected profiles follow the input profiles extremely well
up to the sky threshold ($\sim10$ kpc). The recovered flux within 10
kpc is on average 95\% of the total input flux, with a 1-$\sigma$
spread of 2\%. Recovered effective radii are less accurate, as this
quantity depends quite strongly on the extrapolation of the surface
brightness profile to radii beyond 10 kpc. However, effective radii
derived from the residual-corrected profiles are generally closer to
the true effective radii than those derived from simple S\'ersic fits.

\section{Missing flux in compact quiescent $z\sim2$ galaxies}\label{sec:profiles}

\begin{figure*}
\epsscale{1.1}
\plotone{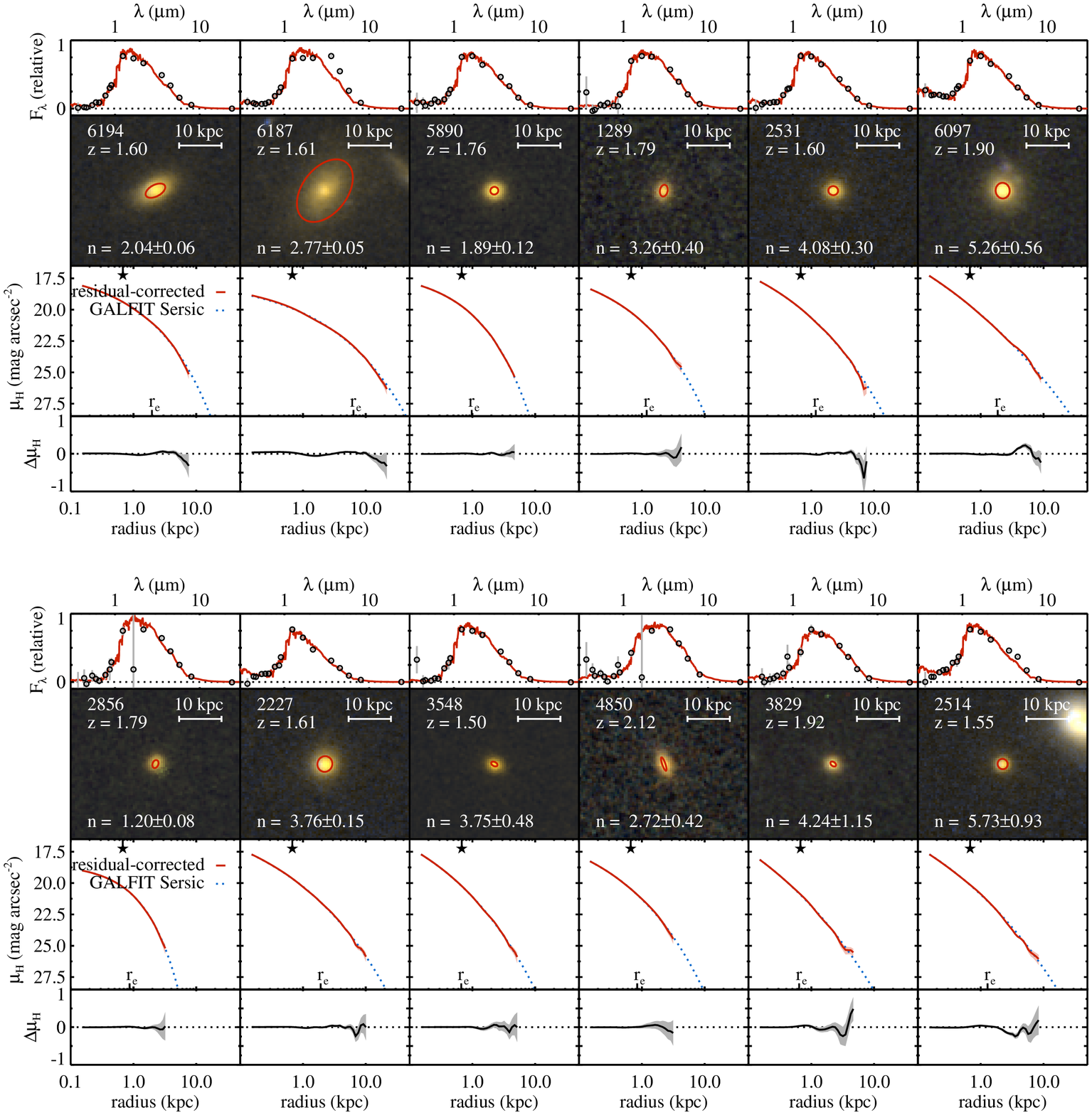}
\caption{Broadband SEDs, color images and PSF-corrected surface
  brightness profiles of $z\sim2$ quiescent galaxies. The SEDs,
  obtained with FAST \citep{kri09}, are based on photometry from the
  FIREWORKS catalog. The color images are composed of rest-frame
  $U_{336}$, $B_{438}$ and $g_{475}$ images, obtained from the
  observed $I_{814}$, $J_{125}$ and $H_{160}$ data. The red ellipses
  are constructed from the best-fitting effective radii, axis ratios,
  and position angles. The best-fit S\'ersic profiles, obtained using
  GALFIT, are indicated by blue dotted curves. Residual-corrected
  surface brightness profiles are shown in red. Effective radii and
  the PSF HWHM are indicated at the bottom and top axes,
  respectively. We are able to measure the true surface brightness
  profiles of these galaxies down to approximately 26 mag
  arcsec$^{-2}$ and out to $r \approx 10$ kpc. In the bottom row we
  show the difference between the best-fit S\'ersic profile and the
  residual-corrected profile. Individual residual-corrected profiles
  show deviations from simple S\'ersic profiles, although these
  deviations are consistent with zero within the
  errors.}\label{fig:profiles}
\end{figure*}
\begin{figure*}
\epsscale{1.1}
\addtocounter{figure}{-1}
\plotone{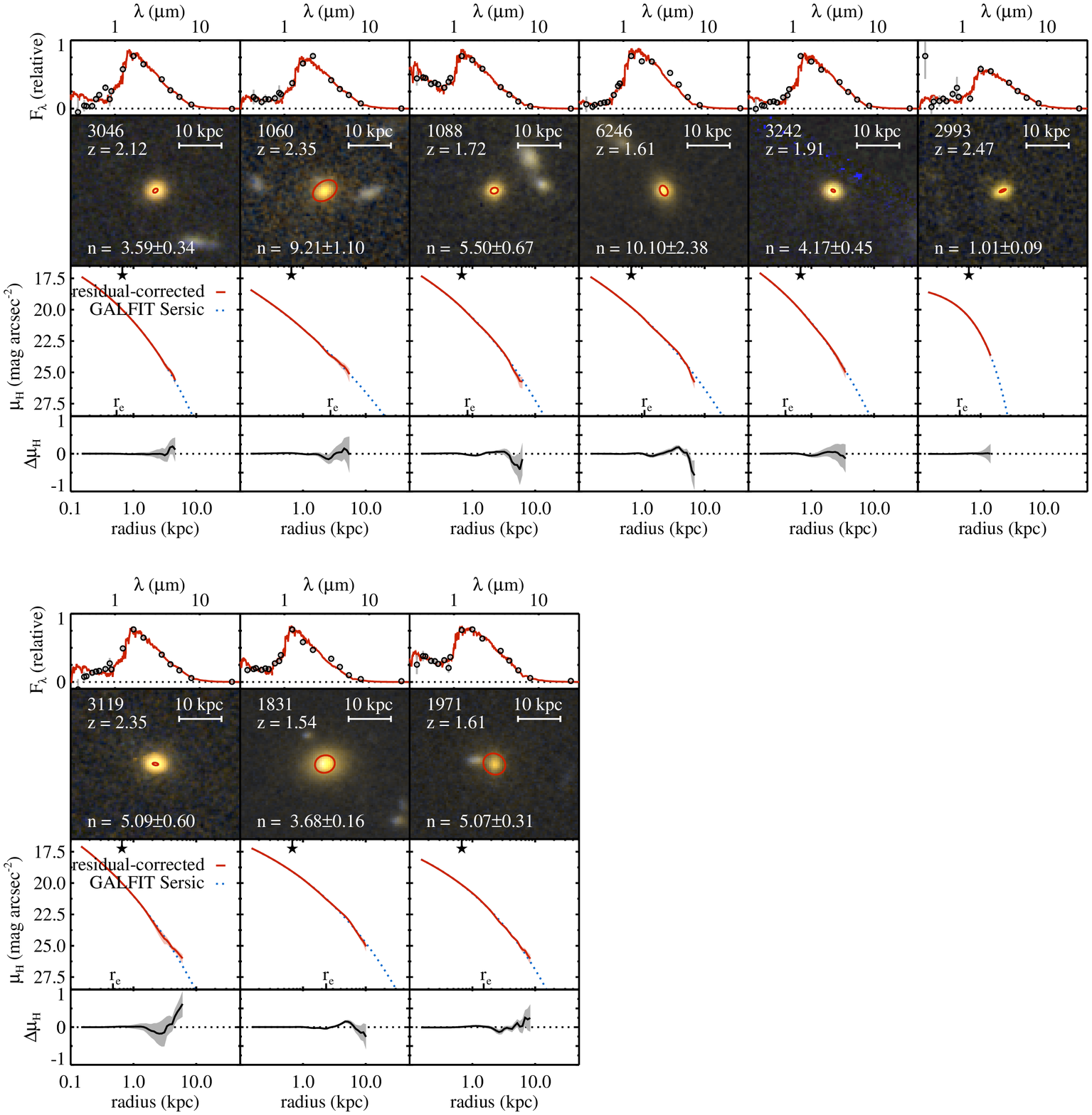}
\caption{Continued from previous page}
\end{figure*}

\begin{table}
\centering
\caption{Surface brightness profiles.\label{tab:profiles}}
\begin{tabular}{c c c c c}
\\
\hline \hline
\colhead{ID$^\mathrm{a}$} & \colhead{$r_\mathrm{arcsec}$} & \colhead{$r_\mathrm{kpc}$} & \colhead{$\mu_H$} & \colhead{$\log \Sigma$} \\
\colhead{} & \colhead{(arcsec)} & \colhead{(kpc)} & \colhead{(AB mag arcsec$^{-2}$)} & \colhead{($\log M_\odot$ kpc$^{-2}$)} \\
\hline
1060 & 0.0180 & 0.147 & $18.413 \pm 0.0010$ & $10.843 \pm 0.0004$ \\
1060 & 0.0198 & 0.162 & $18.548 \pm 0.0011$ & $10.789 \pm 0.0005$ \\
1060 & 0.0216 & 0.177 & $18.673 \pm 0.0013$ & $10.739 \pm 0.0005$ \\
1060 & 0.0240 & 0.196 & $18.826 \pm 0.0015$ & $10.678 \pm 0.0006$ \\
1060 & 0.0264 & 0.216 & $18.966 \pm 0.0019$ & $10.622 \pm 0.0007$ \\
1060 & 0.0288 & 0.235 & $19.095 \pm 0.0021$ & $10.570 \pm 0.0008$ \\
1060 & 0.0318 & 0.260 & $19.244 \pm 0.0024$ & $10.510 \pm 0.0010$ \\
1060 & 0.0348 & 0.285 & $19.382 \pm 0.0027$ & $10.455 \pm 0.0011$ \\
1060 & 0.0384 & 0.314 & $19.534 \pm 0.0032$ & $10.395 \pm 0.0013$ \\
1060 & 0.0426 & 0.348 & $19.696 \pm 0.0037$ & $10.330 \pm 0.0015$ \\
\nodata & \nodata & \nodata & \nodata & \nodata \\
\hline \\
\multicolumn{5}{p{0.45\textwidth}}{{\sc Note}. --- This Table is published in its entirety in the electronic edition of \apj, and can also be downloaded from \url{http://www.strw.leidenuniv.nl/~szomoru/}. A portion is shown here for guidance regarding its form and content.} \\
\multicolumn{5}{l}{$^\mathrm{a}$FIREWORKS ID \citep{wuy08}}
\end{tabular}
\end{table}

We now use the residual-correction method to derive the surface
brightness profiles of the $z\sim2$ quiescent galaxies. The results
are shown in Figure~\ref{fig:profiles}. The SEDs, shown in the top
row, illustrate the low levels of UV and IR emission of the quiescent
galaxies in our sample. Rest-frame color images are shown in the
second row. These images indicate that the galaxies in this sample
generally have compact elliptical morphologies. Some galaxies have a
nearby neighbor; in these cases we simultaneously fit both objects to
account for possible contamination by flux from the companion
object. In the third row, best-fit S\'ersic profiles are shown in blue
and residual-corrected profiles in red. The residual-corrected
profiles follow the S\'ersic profiles remarkably well. Most galaxies
deviate slightly at large radii. The difference between the best-fit
S\'ersic profiles and the residual-corrected profiles are shown in the
bottom row. The deviations are generally small within $2r_e$; for some
galaxies larger deviations occur at larger radii, but in these cases
the uncertainty is very high due to the uncertain sky. Overall, the
profiles are consistent with simple S\'ersic profiles. The profiles
are given in Table~\ref{tab:profiles}, and can also be downloaded from
\url{http://www.strw.leidenuniv.nl/~szomoru/}

\begin{figure}
\epsscale{1.1}
\plotone{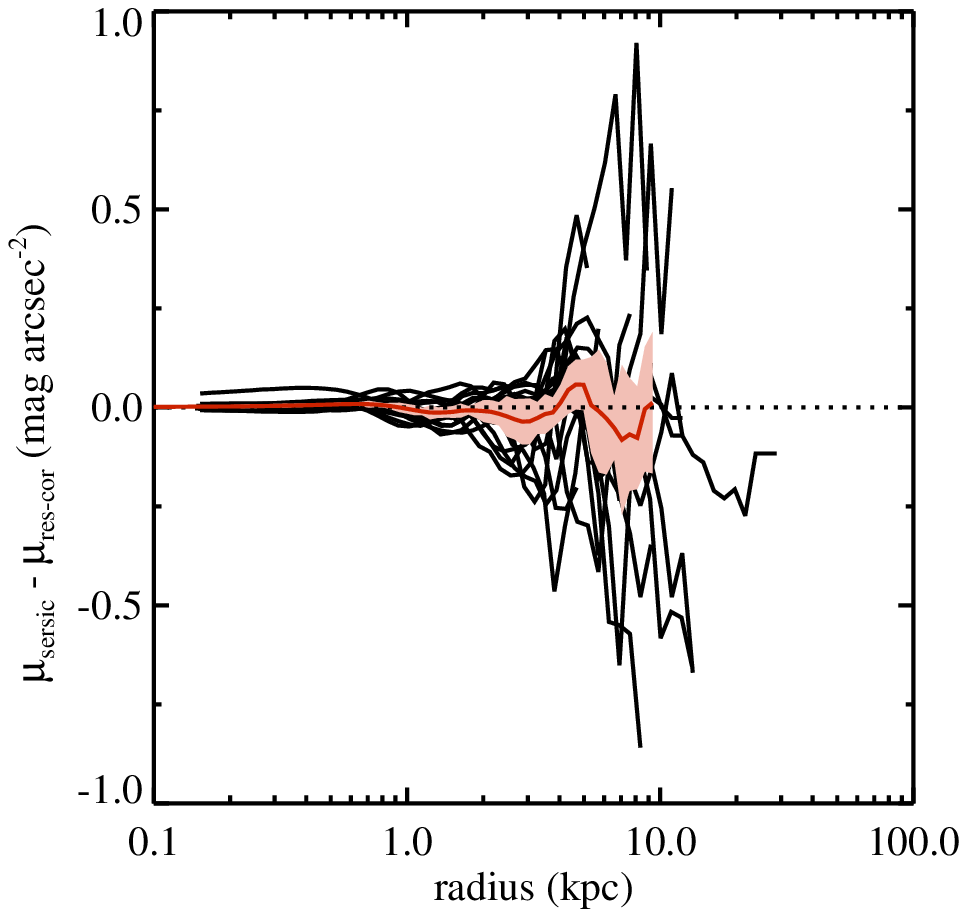}
\caption{Deviations of galaxy profiles from S\'ersic profiles. The
  difference between the best-fit S\'ersic profile and the
  residual-corrected profile is plotted as a function of radius for
  all galaxies in our sample (black lines). The mean profile is shown
  in red, with the shaded light red region indicating the $1-\sigma$
  spread in the distribution. Although individual galaxy profiles
  deviate from S\'ersic profiles, on average the difference is
  consistent with zero.}\label{fig:residuals}
\end{figure}

In order to investigate whether the profiles of $z\sim2$ quiescent
galaxies deviate systematically from S\'ersic profiles we plot the
difference between the best-fit S\'ersic profile and the
residual-corrected flux profile in Figure~\ref{fig:residuals}, for all
galaxies. Black lines indicate the deviation profiles of individual
galaxies, and their mean is indicated by the red line. The light red
area shows the 1-$\sigma$ spread around the mean. The mean profile is
consistent with zero at all radii; the surface brightness profiles of
quiescent galaxies at $z\sim2$ seem to be well described by S\'ersic
profiles. On average the residual correction increases or decreases
the total flux of each galaxy in our sample by only a few percent,
with an upper limit of 7\%. The mean contribution of the residual flux
to the total flux for all galaxies in our sample is -0.7\%. Thus, we
do not find evidence that indicates that there is missing low surface
brightness emission around compact quiescent $z\sim2$ galaxies, and we
therefore conclude that the small sizes found for these galaxies are
correct.

\section{The mass growth of $z\sim2$ quiescent galaxies}

\begin{figure*}
\epsscale{1.1}
\plotone{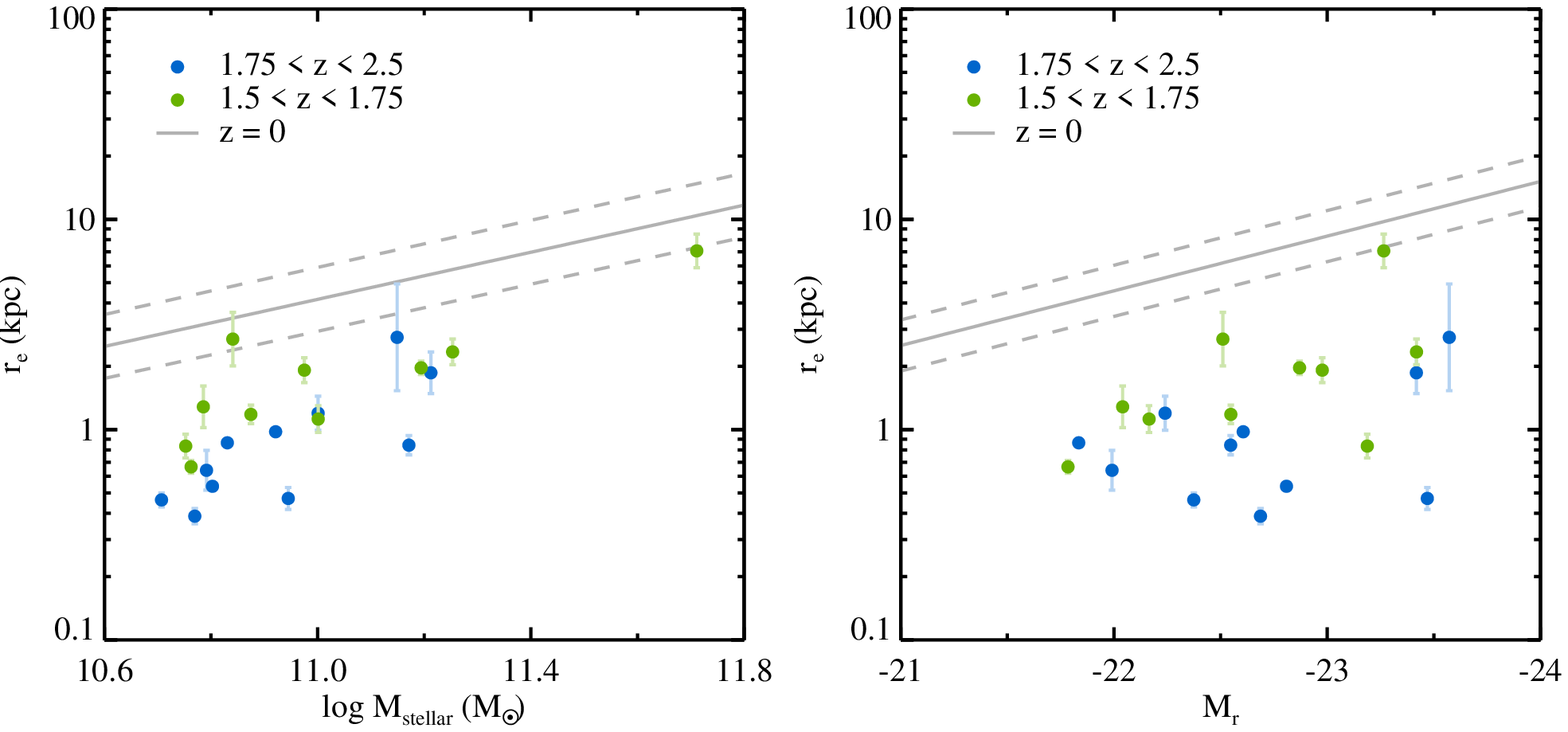}
\caption{Relations between size and stellar mass (left panel) and size
  and rest-frame $r$-band absolute magnitude (right panel). Grey lines
  indicate the low-redshift mass-size and magnitude-size relations
  from \cite{she03}, green and blue points indicate the $z\sim2$
  sample (divided into low and high redshift bins, respectively). The
  $z\sim2$ galaxies are, on average, almost an order of magnitude
  smaller than low-redshift galaxies of similar mass and
  luminosity. However, there is a significant range in sizes at both
  redshifts. The largest $z\sim2$ galaxies lie very close to the $z=0$
  mass-size relation.}\label{fig:masssize}
\end{figure*}

In the previous Section we have shown that the surface brightness
profiles of $z\sim2$ quiescent galaxies closely follow S\'ersic
profiles, and that their sizes are not systematically underestimated
due to a lack of sensitivity. We now compare their size distribution
and surface brightness profiles to those of low-redshift galaxies. In
Figure~\ref{fig:masssize} we show the mass-size and magnitude-size
relations for the $z\sim2$ galaxies and for low-redshift massive
elliptical galaxies, taken from \cite{she03}. The $z\sim2$ sample has
been split into two redshift bins: $1.75 < z < 2.5$ and $1.5 < z <
1.75$ (shown in blue and green, respectively). The low-redshift sample
is shown in grey. Galaxies at $z\sim2$ are significantly smaller than
those at $z=0$. We fit a power law of the form $r_e \propto
(1+z)^\alpha$ and find $\alpha = -0.94\pm0.16$, which is comparable to
e.g., \cite{wel08} and \cite{san11}, but slightly steeper than
\cite{new10} and significantly shallower than \cite{bui08}.

However, the $z\sim2$ galaxies span a large range in size; some are
supercompact, while others are as large as $z=0$ galaxies. Following
\cite{she03}, we quantify this range using $\sigma_{\log_{10} r_e}$,
which is defined as the 1-$\sigma$ spread in $\log_{10} r_e$ around
the median mass-size relation, which we fix to the $z=0$ slope. Note
that we define the scatter in log$_{10}$ basis, not the natural
logarithm as used by \cite{she03}. It is equal to $0.24\pm0.06$ for
our entire sample, while \cite{she03} find values around
$\sigma_{\log_{10} r_e} = 0.16$ for early-type galaxies at $z=0.1$ in
the same mass range. The values for the two high-redshift subsamples
are $0.21\pm0.11$ at $1.5 < z < 1.75$ and $0.19\pm0.07$ at $1.75 < z <
2.5$. These values are upper limits, since they include the errors on
individual size measurements; however, if our error estimates are
correct, their effect on the scatter is $\lesssim 0.01$ dex. The
scatter we measure is comparable to that found in \cite{new11}. These
authors find $\sigma_{\log_{10} r_e} \approx 0.25$ for galaxies with
$10^{10.7} M_\odot < M_{stellar} \lesssim 10^{11.7} M_\odot$ at
$z\sim2$. We note that our sample contains several galaxies that are
part of an overdensity at $z=1.6$ (e.g., \citealt{gil03};
\citealt{cas07}; \citealt{kur09}). In particular, the two largest
galaxies in our sample are part of this overdensity. Excluding the
$z=1.6$ galaxies from our analysis does not significantly alter the
spread in galaxy sizes in the $1.5 < z < 1.75$ redshift bin:
$\sigma_{\log_{10} r_e} = 0.21\pm0.14$.

The size measurements used in \cite{she03} have been shown to suffer
from systematic errors due to background oversubtraction
\citep{guo09}. As a result of this, the mass-size relation measured by
\cite{she03} is significantly shallower than that found by, e.g.,
\cite{guo09}. We therefore repeat our determination of the scatter
around the $z\sim2$ mass-size relation using the \cite{guo09}
measurements. This results in a decrease in the scatter by only
$\sim0.03$ dex, and does not affect our conclusions.

\begin{figure}
\epsscale{1.1}
\plotone{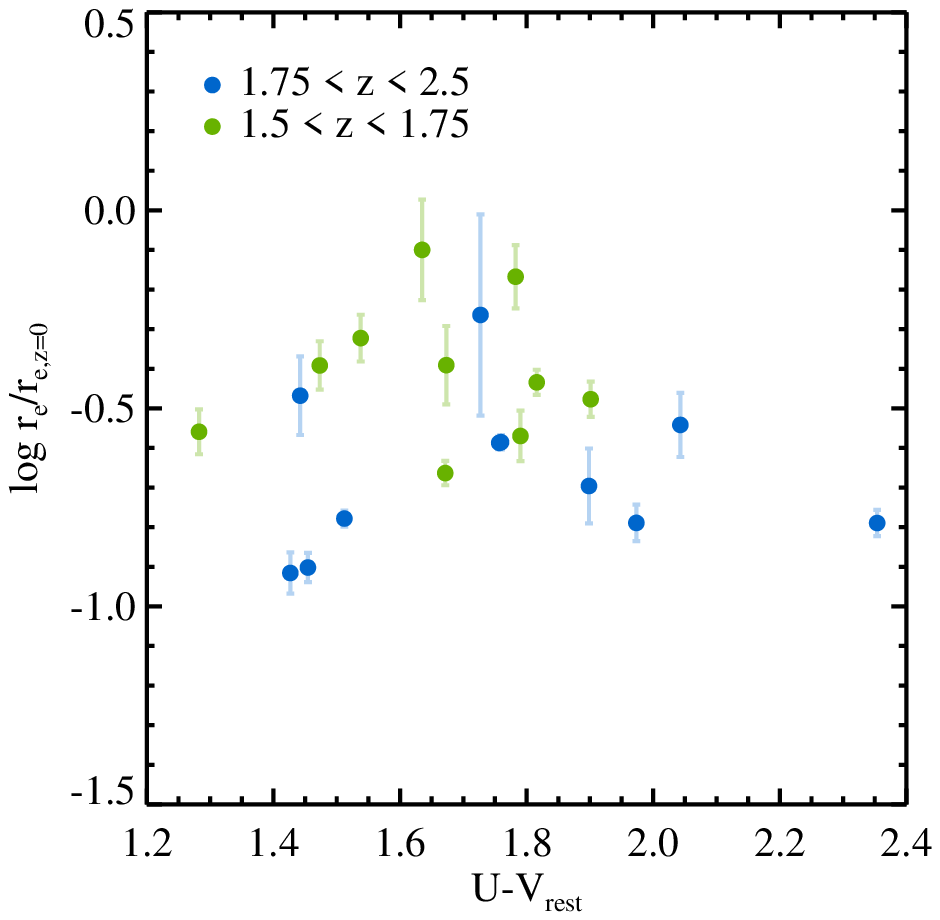}
\caption{Offset from the $z=0$ mass-size relation as a function of
  rest-frame $U-V$ color for galaxies at $1.5 < z < 1.75$ (green
  points) and $1.75 < z < 2.5$ (blue points). The offset is calculated
  by dividing the effective radius of each galaxy by the median
  effective radius of $z=0$ quiescent galaxies with the same mass,
  using the $z=0$ mass-size relation from \cite{she03}. The galaxies
  are split into two redshift bins. Assuming that rest-frame $U-V$
  color is a good proxy for the mean stellar age of galaxies, we find
  no evidence for a correlation between galaxy compactness and galaxy
  age for $z\sim2$ quiescent galaxies.\\}\label{fig:compactness}
\end{figure}

We note that, even within the limited redshift range under
consideration, differences in redshift play a role: the galaxies in
the $1.75 < z < 2.5$ subsample are clearly smaller than the $1.5 < z <
1.75$ galaxies. This may explain some of the disagreement between
studies of high-redshift quiescent galaxies. In particular, the large
effective radii found by \cite{man10} for some high-redshift quiescent
galaxies could be due to the fact that they select galaxies with $1.4
< z < 1.75$. In this context, part of the size evolution between
$z\sim2$ and $z=0$ could be due to the appearance of young, relatively
large quiescent galaxies after $z\sim2$ (e.g., \citealt{dok08};
\citealt{fra08}; \citealt{sar09}; \citealt{wel09};
\citealt{cas11}). We note that \cite{sar09} find evidence for a
correlation of galaxy compactness with stellar age, such that the most
compact high-redshift quiescent galaxies contain older stellar
populations than quiescent galaxies that lie close to the $z=0$
mass-size relation. We investigate this correlation in
Figure~\ref{fig:compactness}, using rest-frame $U-V$ color as a proxy
for galaxy age. We define galaxy compactness as the offset between the
$z\sim2$ galaxy sizes and the $z=0$ mass-size relation of
\cite{she03}: $r_e/r_{e,z=0} = r_e/(2.88 \times 10^{-6} \times
M^{0.56})$. We find no evidence for a correlation between galaxy
compactness and galaxy age in our data.

\begin{figure}
\epsscale{1.1}
\plotone{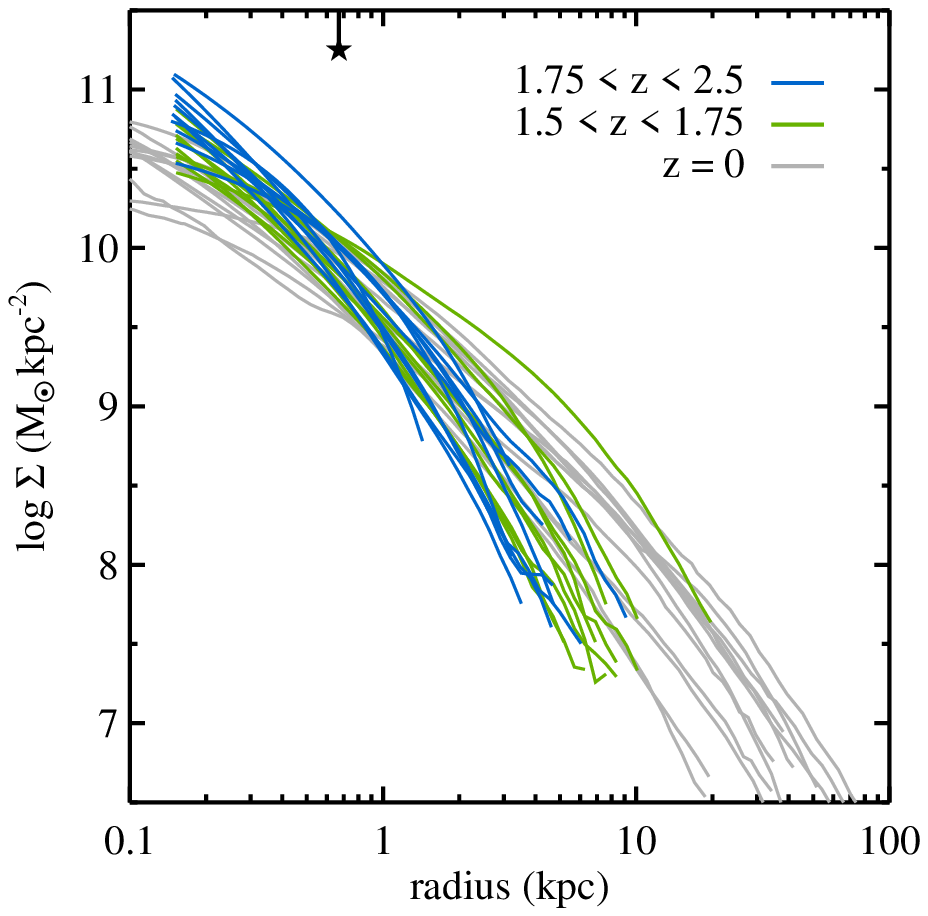}
\caption{Comparison of stellar mass surface density profiles of
  $z\sim2$ galaxies (blue and green curves) to elliptical galaxies in
  the Virgo cluster (\citealt{kor09}; grey curves). The Virgo galaxies
  are selected to have masses equal to or higher than those of the
  high-redshift galaxies. Radial color gradients are ignored when
  calculating the mass density profiles. The star, top left, indicates
  the PSF HWHM at $z=2$. The central densities of the $z\sim2$
  galaxies are very similar to those of the $z=0$ galaxies. At larger
  radii, however, significant evolution must occur if the $z\sim2$
  galaxies are to evolve into massive low-redshift elliptical
  galaxies.\\}\label{fig:virgoprofiles}
\end{figure}

In Figure~\ref{fig:virgoprofiles} we compare the stellar mass surface
density profiles of the $z\sim2$ galaxies to those of low redshift
galaxies. Based on their masses and number densities, we expect
$z\sim2$ quiescent galaxies to evolve into the most massive
low-redshift galaxies (e.g., \citealt{dok10}). As a comparison sample
we therefore use surface brightness profiles of elliptical galaxies
with equal or higher mass in the Virgo cluster from
\cite{kor09}. These authors used a combination of space-based and
ground-based observations to obtain surface brightness profiles with
very high resolution and dynamic range, covering almost three orders
of magnitude in radius. The surface brightness profiles have been
converted to stellar mass surface density profiles using the total
stellar mass-to-light ratios. We have ignored radial color gradients,
which are known to exist at low and high redshift (e.g., \cite{dok10};
\cite{szo11}; \cite{guo11}). These profiles are shown in grey, with
the profiles of the $z\sim2$ galaxies overplotted in blue and green.

What is most apparent in Figure~\ref{fig:virgoprofiles} is that the
central ($r < 1-3$ kpc) surface densities of the $z\sim2$ galaxies are
very similar to those of the $z=0$ galaxies, while at larger physical
radii (in kpc) the high-redshift galaxies have lower surface densities
than the low-redshift galaxies. The profiles are in close agreement
with previous studies (e.g., \citealt{bez09}; \citealt{car10}). We
compare the change in radial mass density profiles to the mass
evolution of quiescent galaxies described in \cite{bra11}. These
authors show that galaxies with a number density of $10^{-4}$
Mpc$^{-3}$ have grown in mass by a factor $\sim2$ since $z=2$. As
mentioned above, the mass contained within 3 kpc changes very little
from $z\sim2$ to $z=0$; we find an increase on the order of
10\%. However, the mass contained outside 3 kpc is approximately ten
times higher for the $z=0$ galaxies than for the $z\sim2$ galaxies,
and is equal to 58\% of their total mass. Thus, slightly more than
half of the total mass of the $z=0$ ellipticals is located at $r > 3$
kpc, whereas the $z\sim2$ galaxies contain nearly no mass at these
radii. This is consistent with the \cite{bra11} result, and suggests
that compact $z\sim2$ quiescent galaxies may survive intact as the
cores of present-day massive ellipticals, with the bulk of mass
accretion since $z\sim2$ occuring at large radii. This is consistent
with an inside-out scenario of galaxy growth, as described in e.g.,
\cite{dok10}. We note that this discussion ignores transformations of
star forming galaxies to the quiescent population.

\begin{figure*}
\epsscale{1.1}
\plotone{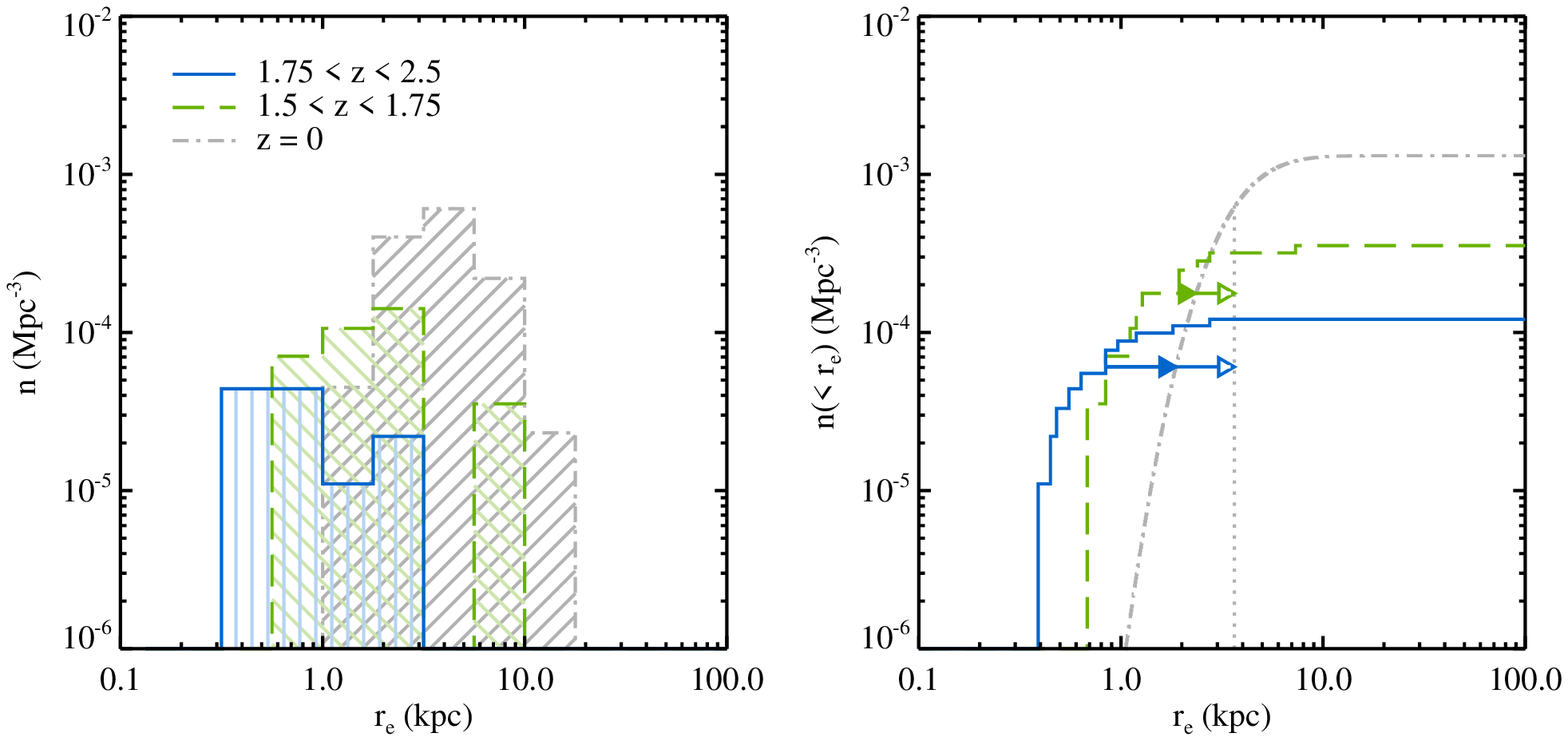}
\caption{Comoving number density (left panel) and cumulative comoving
  number density (right panel) as a function of effective radius
  $r_e$, for galaxies at $1.75 < z < 2.5$, $1.5 < z < 1.75$, and $z=0$
  (solid blue, dashed green, and dot-dashed grey lines,
  respectively). The $z=0$ number densities are obtained by combining
  the stellar mass function of \cite{bel03} with the mass-size
  relation of \cite{she03}. The $z~2$ number densities have been
  scaled such that the total number density corresponds to the results
  of \cite{bra11}. Both the median effective radius and the total
  number density of quiescent galaxies show a strong increase from
  $z\sim2$ to $z=0$. The solid arrows in the right-hand panel indicate
  the minimum size growth required for high-redshift galaxies to grow
  into the smallest galaxies at $z=0$. The open arrows indicate the
  size growth required for high-redshift galaxies to grow to the same
  median size as galaxies at $z=0$. The minimum size growth required
  for $z\sim2$ quiescent galaxies is approximately a factor 2 smaller
  than the median size growth between $z\sim2$ and
  $z=0$.}\label{fig:numdens}
\end{figure*}

Finally, we compare the comoving number densities and comoving
cumulative number densities of our $z\sim2$ sample to the number
densities of $z=0$ galaxies in Figure~\ref{fig:numdens}. To obtain the
$z=0$ number densities we combine the $z=0$ mass function for
early-type galaxies from \cite{bel03} with the mass-size relation of
\cite{she03}: we use the relations appropriate for early-type galaxies
and evaluate over the mass range $5\times10^{10} M_\odot < M_{stellar}
< 5\times10^{11} M_\odot$. Given our small field size we cannot
determine number densities accurately. We therefore adopt the number
densities measured by \cite{bra11}. These authors used data covering a
much larger field of view (approximately 25 times larger than the
CANDELS GOODS-South field), and as such their results are less
sensitive to cosmic variance. We scale our (cumulative) number density
distributions such that the total number density corresponds to the
\cite{bra11} results. We note that our measured number densities are
approximately a factor 2 smaller than those in \cite{bra11},
consistent with expectations from field-to-field variations
\citep{som04}. We first consider the comoving number density
distributions, plotted in the left panel of
Figure~\ref{fig:numdens}. As expected, the median radius and the total
number density increase with time, as existing galaxies grow in size
and new quiescent galaxies appear. $r_{e,median} = 0.84\pm0.20$ kpc,
$1.92\pm0.45$ kpc, and $3.82\pm0.03$ kpc at $1.75 < z < 2.5$, $1.5 < z
< 1.75$ and $z=0$, respectively.

We can place constraints on the minimum size growth of $z\sim2$
galaxies by considering comoving cumulative number densities, shown in
the right-hand panel of Figure~\ref{fig:numdens}. We assume that the
population of $z\sim2$ quiescent galaxies grows just enough to fall
within the $z=0$ size distribution, but doesn't necessarily grow to
the same median size as $z=0$. This results in a shift of the $z\sim2$
cumulative number density distribution, indicated by the filled arrows
in Figure~\ref{fig:numdens}. This shift is approximately a factor
$\sim2$ smaller than the size growth required for the $z\sim2$
quiescent galaxy population to match the median size at $z=0$
(indicated by the open arrows). Thus, in this minimal-growth scenario,
half of the observed size evolution between $z\sim2$ and $z=0$ is due
to the growth of existing galaxies, while the other half results from
the appearance of new, larger quiescent galaxies at intermediate
redshifts. These results are consistent with e.g., \cite{cas11} and
\cite{new11}.

\section{Summary and conclusions}

In this Paper we have demonstrated that the small measured sizes of
$z\sim2$ massive quiescent galaxies are not caused by a lack of
sensitivity to low surface brightness flux. Using deep data and a
method which is sensitive to excess emission at large radii, we have
shown that the surface brightness profiles of these galaxies are well
described by S\'ersic profiles. The median S\'ersic index is
$n_{median} = 3.7$, similar to low-redshift quiescent galaxies.

The sizes of $z\sim2$ quiescent galaxies span a large range; although
the median effective radius is small ($r_{e,median} = 1.1$ kpc),
values up to $\sim7$ kpc are observed. The scatter in $\log r_e$ is
0.24 at $z\sim2$, aproximately 1.5 times as large as at $z=0$. This
indicates that the ``dead'' population of galaxies is very diverse at
$z\sim2$. We note that the size evolution between $z=1.5$ and $z=2.5$
is significant, which suggests that the cause of discrepancies in the
results of different studies of the measured sizes of quiescent
galaxies around $z=2$ could be due to small differences in the
redshift ranges considered.

Additionally, we have compared the stellar mass surface density
profiles of $z\sim2$ galaxies to those of massive early-type galaxies
in the Virgo cluster. Although the densities within $\sim1$ kpc are
comparable, at larger radii the $z\sim2$ galaxies show a clear deficit
of mass. This puts strong constraints on models of galaxy formation
and evolution. Firstly, most of the size buildup of $z\sim2$ quiescent
galaxies must occur at large radii ($> 1$ kpc). Secondly, a
significant contribution from major gas-rich mergers since $z\sim2$
seems to be ruled out, as this would disturb the inner density
profiles of these galaxies. Minor, dry merging and slow accretion of
matter seems to be the most viable method of evolving these galaxies
into their $z=0$ descendants.

Finally, we have investigated the evolution in the size distribution
of massive quiescent galaxies. We conclude that the median size of
massive quiescent galaxies changes by a factor $\sim4$ between
$z\sim2$ and $z=0$, and is accompanied by an increase in number
density of a factor $\sim7$. However, it is important to note that the
size growth of individual galaxies is likely to be significantly
smaller. The minimum required size growth for the $z\sim2$ quiescent
galaxy population is approximately a factor $\sim2$ smaller than the
median overall size growth. In this scenario the stronger overall size
growth may be caused by the appearance of new, larger quiescent
galaxies at intermediate redshifts.

One of the main observational uncertainties pertaining to the size
evolution of massive quiescent galaxies now appears to be resolved;
robust sizes, measured at high resolution and using very deep
rest-frame optical data, indicate that galaxies at $z\sim2$ were
significantly smaller than equally massive galaxies at $z=0$. However,
the mechanisms driving this evolution and their precise effects on the
structure of individual galaxies, as well as on the characteristics of
the population as a whole, are still not entirely understood. Most
studies seem to point towards gas-poor galaxy merging as the dominant
growth process (e.g., \citealt{bez09}; \citealt{naa09};
\citealt{hop10}); however, it is unclear whether this can account for
all the observed size growth. A complicating factor in such studies is
that tracing the same group of galaxies across cosmic time is very
difficult, since their masses, sizes and stellar population properties
are not constant; selecting the same population of galaxies at
different epochs is therefore not trivial. Studies at fixed
(cumulative) number density may provide a solution to this problem,
though only for relatively massive galaxies.

\acknowledgments

This work is based on observations taken by the CANDELS Multi-Cycle
Treasury Program with the NASA/ESA HST, which is operated by the
Association of Universities for Research in Astronomy, Inc., under
NASA contract NAS5-26555.

{\it Facilities:} \facility{HST (ACS, WFC3)}.

\end{document}